\documentclass{aa}
\usepackage{amssymb}
\usepackage{graphicx}
\usepackage{longtable}
\usepackage{xcolor}
\usepackage{url} 
\usepackage{amsmath}

\def\Teff{$T_{\rm eff}$}
\def\logg{$\log\,g$}
\def\Vt{V${\rm t}$}
\newcommand{\kms}{km~s$^{-1}$}

\def\apjl{ApJ}%
\def\apjs{ApJS}%
\def\ao{Appl.~Opt.}%
\def\aap{A\&A}%
\begin{document} 

\title{Molybdenum and ruthenium in the Galactic disk: A closer look 
at their nucleosynthesis components}

\author{T.~Mishenina\inst{1},
                   T.~Kurtukian-Nieto\inst{2},               
          T.~Gorbaneva\inst{1},
         A. M.~Amarsi\inst{3},
          A.~Psaltis\inst{4}\thanks{The NuGrid collaboration, http://www.nugridstars.org},
           M.~Pignatari\inst{5,6,7}\thanks{The NuGrid collaboration, http://www.nugridstars.org}
          }
    
\authorrunning{T.~Mishenina  et al.}

\institute
	{Astronomical Observatory, Odesa National University, 1b, Marazlievska str., 65014, Odesa, Ukraine\\ \email{tmishenina@ukr.net}\\              
        \and Instituto de Estructura de la Materia, IEM-CSIC, Serrano 121, 28006 Madrid, Spain  \\
        \and Theoretical Astrophysics, Department of Physics and Astronomy, Uppsala University, Box 516, SE-751 20 Uppsala, Sweden \\
        \and Department of Astronomy \& Physics, Saint Mary’s University, Halifax, NS B3H 3C3, Canada
        \\
        \and Konkoly Observatory, HUN-REN, Konkoly Thege Miklos ut 15-17, H-1121 Budapest, Hungary\\ \email{mpignatari@gmail.com}\\
        \and MTA Centre of Excellence, Budapest, Konkoly Thege Miklós út 15-17, H-1121, Hungary\\
        \and University of Bayreuth, BGI, Universitätsstraße 30, 95447 Bayreuth, Germany                }

\date{Received September 15, 1996; accepted March 16, 1997}

\abstract
{
The stellar origin of the elements molybdenum (Mo, Z=42) and ruthenium (Ru, Z=44) is still a matter of debate.  
Studying their abundances provides valuable insights into nucleosynthesis processes and the broader evolution of neutron-capture elements.} 
{The aim of this work is to present 
new observations of Mo and Ru, together with nearby neutron-capture elements strontium (Sr) and zirconium (Zr) for a new sample of 154 giant stars, located in the Galactic disk with metallicities -1 < [Fe/H] < +0.3. }
{
The abundances were determined under the assumption of the local thermodynamic equilibrium by ﬁtting synthetic spectra. The abundances of Mo were derived from the Mo I lines at 5506 and 5533 \AA,
while the abundances of Ru were determined from Ru I lines at 4584, and 4757 \AA.  
The abundances of Sr and Zr were
measured using lines of Sr II at 4077 and 4215 \AA~and lines of Zr I at 5385, 6127, and 6134 \AA.}
{We derived Sr, Zr, and Mo abundances for all stars in our sample, and Ru for 110 stars. 
For most of the giant stars observed in this work, Mo and Ru abundances were determined for the first time. We compare our observations with the signatures from different nucleosynthesis processes.} 
{Both the [Mo/Fe] and [Ru/Fe] in our stars show a decreasing trend with respect to increasing [Fe/H]. This pattern is similar to that of [Zr/Fe], whereas [Sr/Fe] exhibits a relatively flat trend with metallicity. These results are compatible with our previous measurements obtained for dwarf stars over the same metallicity range. Compared to the $s$-process ratios, all stars show a lower [Zr/Mo] and a higher [Ru/Mo], as expected from classical nucleosynthesis. 
Still, it is unclear if additional contributions from neutrino-wind components or the $i$-process is needed to explain the observed scatter of [Zr/Mo] and [Ru/Mo] in the Milky Way disk. Indeed, such a dispersion is consistent with the variations also seen in r-II stars at low metallicity and could therefore result from the combined contributions of $r$-process and $s$-process to galactic chemical evolution. The observed [Zr/Mo] and [Ru/Mo] scatter in r-II stars should be constrained by future investigations to define if any contributions of additional nucleosynthesis components are needed.   
} 


\titlerunning{Molybdenum and Ruthenium in the Galactic Disk}
\maketitle

\section{Introduction}

The elements molybdenum (Mo, Z = 42)  and ruthenium (Ru, Z = 44) are 
composed of several stable isotopes with distinct nucleosynthetic origins, and studying their abundances provides valuable insights into the  
galactic chemical evolution (GCE) of neutron-capture elements \citep[e.g.,][]{travaglio:04, bisterzo:14, kobayashi:20, prantzos:20}. 
$^{92}$Mo, $^{94}$Mo, $^{96}$Ru, and $^{98}$Ru can be produced by the $\gamma$-process or by proton-rich neutrino wind components \citep[][]{arnould:03, rauscher:13, pignatari:16a}. $^{96}$Mo and $^{98}$Ru are defined as $s$-only isotopes, as they are made by the $s$-process; however, they are shielded by stable isobars from any contribution of other neutron-capture processes \citep[e.g.,][]{kaeppeler:11, lugaro:23}. $^{100}$Mo and $^{104}$Ru are $r$-only isotopes, as the half-lives of the unstable isotopes $^{99}$Mo and $^{103}$Ru are too short to efficiently activate an $s$-process neutron-capture flow toward them \citep[e.g.,][]{cowan:21}. The remaining stable isotopes of Mo ($^{95}$Mo, $^{97}$Mo, and $^{98}$Mo) and Ru ($^{99}$Ru, $^{101}$Ru, and $^{102}$Ru) can be produced by both the $s$-process and the $r$-process. Additional nucleosynthesis components have been proposed as sources of Mo and Ru for stellar archaeology and/or for GCE, including neutrino-driven ejecta from core-collapse supernovae \citep[e.g.,][]{montes:07, arcones:11, farouqi:10, bliss:18, psaltis:24} and the intermediate neutron-capture process ($i$-process) \citep[e.g.,][]{roederer:16, banerjee:18, cote:18a, choplin:24}.

\cite{peterson:11}, for example, discussed the high overabundances of Mo observed in metal-poor giants such as HD 94028 and HD 160617. 
They suggested that Mo could be produced in the low-entropy regime of a high-entropy wind (HEW) above the neutron star formed during a Type II supernova \citep[e.g.,][]{farouqi:09}. Subsequently, \cite{Hansen:14} conducted a detailed analysis of Mo enrichment in metal-poor stars and revealed considerable scatter in Mo abundances in both dwarfs and giants. This scatter indicated contributions from multiple formation processes beyond HEWs. 

In the Galactic disk, GCE models, such as those by \cite{prantzos:18}, \cite{mishenina:19a}, and \cite{kobayashi:20}, which incorporated contributions from multiple stellar sources such as low- and intermediate-mass (LIM) stars,
massive stars, and fiducial $r$-process sources, were unable to fully reproduce the observed trends of Mo and Ru. 
Concerning solar composition, \cite{travaglio:04}, \cite{bisterzo:14}, and \cite{prantzos:20} derived the total $s$-process contribution to be 39$\%$ and 24$\%$, 39$\%$ and 29$\%$, and 50$\%$ and 34$\%$, respectively. 
The $r$-process contribution may vary depending on the $r$-process setup and the reference composition adopted in GCE simulations, leaving open the possibility to a contribution from other nucleosynthesis processes besides the classical $s$-process and $r$-process. 
\cite{travaglio:04} found that for elements between Sr and Ba (including Mo and Ru), an additional $s$-process component was needed to reproduce the solar abundances, introducing the lighter-element primary process or LEPP. The existence and properties of LEPP remain an open question \citep[e.g.,][and references therein]{montes:07, cristallo:15, prantzos:20}. In particular, the LEPP component was originally associated with the signatures observed in several metal-poor stars showing high production at the Sr peak and marginal or no production of the lanthanides; that is, it was not produced together with $r$-process elements \citep[][]{montes:07}.
As mentioned earlier, additional contributions to Mo and Ru (if needed) could be also due to other sources, such as the $i$-process or neutrino-driven ejecta components from core-collapse supernovae (CCSNe). 

In \cite{mishenina:19a}, we provided a large sample of Mo and Ru observations for F-, G-, and K-stars in the Milky Way disk, discussing their potential astrophysical sources. Further studies, such as \cite{forsberg:22}, focused on Mo abundances in thin and thick disks as well as bulge giants, revealing a higher Mo abundance in the bulge compared to the disk. 
More recently, \cite{vanderswaelmen:23} studied Mo abundance together with Pr, Nd, and Eu 
in the Galactic thin disk population, highlighting the difficulty of reproducing Mo using classical nucleosynthesis stellar sources.    
\cite{molero:23} also investigated Mo, Pr, and Nd abundance patterns at different radial gradients in the Galactic thin disk, finding 
good agreement with our results in \cite{mishenina:19a}.  
Finally, \cite{magrini:23} used the data from 62 open clusters to study the Galactic radial abundance 
of $s$-process and $r$-process elements (including Mo).  

In this work, we provide Mo and Ru abundances for a new sample of thin disk giants,
thereby expanding the available observational dataset. 
We conduct an in-depth analysis of the sources of Mo and Ru production by utilizing the latest nucleosynthesis calculations, including contributions from the $s$-process, the $i$-process, 
and neutrino-driven wind nucleosynthesis. 
 
\par
The paper is organized as follows. 
The observations and parameters of stars are described in 
\S \ref{sec: stellar target}.
The abundance determinations and the error analysis are presented in 
\S \ref{sec: abundance determination}. 
Results and nucleosynthesis interpretation are reported in \S \ref{sec: results} and \S \ref{sec: discussion}, and the summary and conclusion are provided in \S \ref{sec: conclusions}.

\section{Observations and atmospheric parameters}
\label{sec: stellar target}

The spectra of the studied stars were obtained using the 1.93~m telescope at the Observatoire de Haute-Provence (France) with the ELODIE echelle spectrograph \citep{baranne:96}, which has a resolution power of R=42,000 and covers the wavelength range 4400–6800 \AA. The signal-to-noise ratio (S/N) was approximately 130–230 at 5500 \AA. Primary processing of the spectra, including image extraction, cosmic ray removal, and flat-fielding, followed the procedures described in \cite{katz:98}.
Further spectral analysis, such as continuum normalization and equivalent width (EW) measurements, was performed using the DECH30 software package developed by G. Galazutdinov \footnote{http://gazinur.com/DECH-software.html}.
\par
For this study, we adopted the atmospheric parameters of giant stars determined in a previous analysis by \cite{mishenina:06}. Briefly, the effective temperature (\Teff) was derived using a set of 100 line-ratio--\Teff\  relations from \cite{kovtyukh:06}. Surface gravity (\logg) was determined using two methods. The first method, the iron ionization equilibrium method, ensures identical average iron abundances from the neutral iron Fe~I and ionized iron Fe~II lines. The second method, the Ca~I wing fitting method, provides detailed fitting of the wings of the Ca~{\sc i} 6162 \AA~ line, as described in \citep{cayrel:96}.
The microturbulent velocity (\Vt) was determined by requiring the iron abundance derived from Fe~{\sc i}  lines to be independent of the EW line. The metallicity ([Fe/H]) was taken as the iron abundance derived from Fe~{\sc i} lines. For a detailed description of the determination of atmospheric parameters and comparisons with the results of other studies, see \cite{mishenina:06}.
For stars also analyzed in other recent studies, we compared our atmospheric parameter values and results with those of \cite{tautvai:21} for disk giants (Zr abundances only, as Mo abundance were not determined), \cite{forsberg:19} (Zr abundances), and \cite{forsberg:22} (Mo abundances).  
In particular, common stars with the study of \cite{tautvai:21} are HD 176408,  HD 166578,  HD 176598, HD 168653, HD 161178, HD 180711, and HD 188119. Common stars with the studies of \cite{forsberg:19} and \cite{forsberg:22} are HD16400, HD 61935, HD 146388, HD 152224, HD 156874, HD 176408, HD 180711, and HD 215030. The mean differences are given in Table \ref{compar}.

\begin{table*}
\caption{Comparison of stellar parameters with other studies for stars common to our sample.}
\label{compar}
\begin{tabular}{llcccccc}
\hline
Reference  & $\Delta$ \Teff  & $\Delta$ \logg & $\Delta$ \Vt & $\Delta$ [Fe/H] & $\Delta$ [Zr/Fe] & $\Delta$ [Mo/Fe] &  N \\
\hline
- &  $\sigma$,$\pm$   & $\sigma$, $\pm$ & $\sigma$, $\pm$ & $\sigma$, $\pm$ & $\sigma$,$\pm$      &  $\sigma$,$\pm$      & \\
\hline 
\cite{tautvai:21}     &   -8.24 &   -0.23&   ---&  0.00& -0.06 &--- &              7 \\
-	& 51.87&     0.22&     &    0.08&  0.09 &  &      \\
\hline
\cite{forsberg:22}  &   -24.75& -0.31& -0.02&  0.02& 0.04 &  0.06& 8 \\
-	& 40.55 &     0.09 &     0.11 &    0.04 &  0.03&  0.04  &      \\
\hline                                                                                 
\end{tabular}
\end{table*}

The small differences observed in metallicity ([Fe/H]) and molybdenum abundance ([Mo/Fe]) between our results and those of \cite{forsberg:22} further validate the reliability of our parameter determinations. These comparisons suggest a consistent and robust determination of stellar parameters across studies.
The total errors in the \logg\  determination for giants are approximately 0.2–0.3 dex, as reported by \cite{mishenina:06}. Comparisons of our atmospheric parameter determinations with those from  \cite{tautvai:21} and \cite{forsberg:22} indicate the following accuracies for our parameters: $\Delta$\Teff = $\pm100~K$, $\Delta$\logg = $\pm0.3$, and $\Delta$\Vt = $\pm0.1$ \kms.

\section{Determination of the elemental abundances}
\label{sec: abundance determination}

\begin{figure}
\begin{tabular}{c}
\includegraphics[width=8cm]{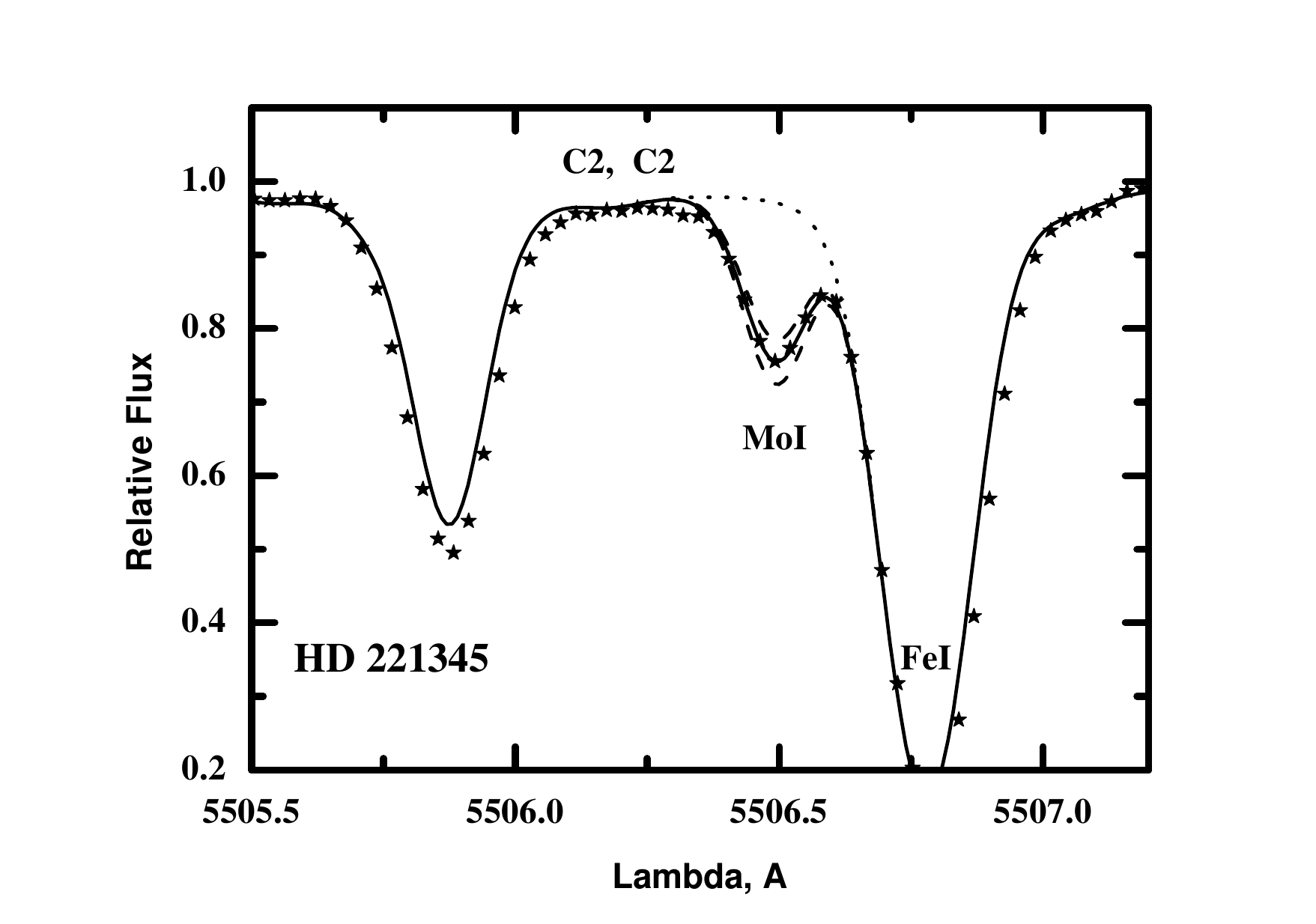}\\
\includegraphics[width=8cm]{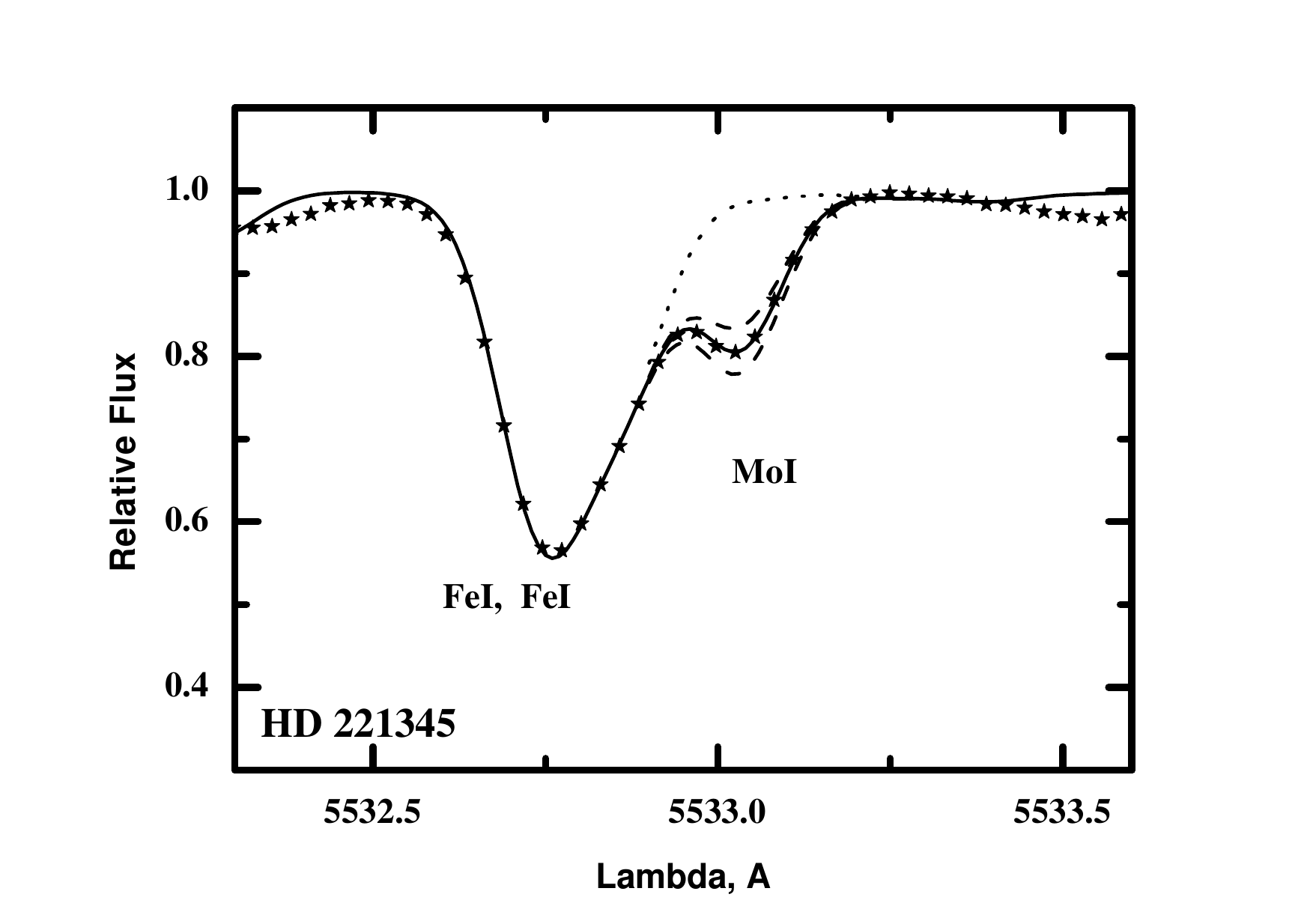}\\
\end{tabular}
\caption{HD 221345: Synthetic (solid line) and observed (asterisks) spectra for Mo I lines. Spectra calculated with a Mo abundance differing by 0.1 dex and without Mo are shown as dashed and dotted lines, respectively.}
\label{mo_prof}
\end{figure}

The elemental abundances were derived using the local thermodynamic equilibrium (LTE) approach, employing the models of \cite{castelli:04} and the modified STARSP LTE spectral synthesis code \citep{tsymbal:96}.
For the Mo I lines at 5506 and 5533 \AA, and the Ru I lines at 4584 and 4757 \AA, oscillator strengths (log gf) were adopted from the latest version of the VALD database \citep{kupka:99}. Figure \ref{mo_prof} shows examples of synthetic and observed spectral fits for the Mo I lines in the star HD 221345, illustrating the quality of the spectral synthesis.

In addition to Mo and Ru, the abundances of strontium (Sr) and zirconium (Zr) were determined. For Sr, we analyzed the Sr II lines at 4077.71 \AA~ and 4215.52 \AA. For the Sr II line at 4077.71 \AA, the atomic data (log gf and HFS) were taken from \citep{bergemann:12}. For Zr, the Zr I lines at 5385.15, 6127.47, and 6134.58 \AA~were used. As with the 4215.52 \AA~ Sr II line, the log gf values were adopted from the latest version of the VALD database \citep{kupka:99}. Table \ref{atomic} gives the wavelength ($\lambda$, \AA), low-level potential (E$_{low}$), and oscillator strength (log gf) of the lines used.

\begin{table}
\caption{Atomic data of used lines.}
\label{atomic}
\begin{tabular}{llcc}
\hline
Species &       $\lambda$, \AA &     E$_{low}$   &          log gf \\ 
\hline
Sr II     &    4077.71    &      0.00       &        0.158   \\
Sr II     &    4215.52   &      0.00       &        -0.155   \\
Zr I    &      5385.15    &      0.519   &         -0.709  \\
Zr I    &       6127.44   &     0.154     &       -1.059   \\
Zr I     &      6134.58    &      0.00    &         -1.279  \\
Mo I   &      5506.49     &     1.335    &        0.060  \\
Mo I  &       5533.03    &     1.335    &        -0.068   \\
Ru I    &     4584.44   &     1.002      &       -0.549   \\
Ru I     &     4757.76   &      1.125    &        -2.666   \\
\hline                                                                                  
\end{tabular}
\end{table}

\subsection{Solar abundances and error estimation}

The adopted LTE solar abundances for Sr, Zr, Mo, and Ru are log A(Sr)$_\odot$ = 2.87 $\pm$ 0.07, log A(Zr)$_\odot$  = 2.58 $\pm$ 0.04, log A(Mo)$_\odot$ = 1.88$\pm$0.08, and log A(Ru)$_\odot$ = 1.75$\pm$0.08, as reported by \cite{asplund:09}.
To estimate possible systematic errors in the Sr, Zr, Mo, and Ru abundance determinations, we analyzed the impact of uncertainties on atmospheric parameters (\Teff/\logg/[Fe/H]/\Vt). This analysis was performed for two representative stars: HD 4482 (4917/2.65/0.02/1.4) and HD 120420 (4676/2.15/-0.27/1.25). We computed the abundances of Sr, Zr, Mo, and Ru for models with modified atmospheric parameters, using the following variations: $\Delta$\Teff = $\pm100~K$, $\Delta$\logg = $\pm0.3$, and $\Delta$\Vt = $\pm0.1$ \kms. See Table \ref{errors}.  
\begin{table*}
\caption{Influence of parameter uncertainties on abundance determination.}
\label{errors}
\begin{tabular}{llcccccccc}
\hline
& & HD4482  && & &HD120240 && & \\
 AN & El  & $\Delta$ \Teff+  & $\Delta$ \logg+ & $\Delta$ \Vt+ & tot+ &$\Delta$ \Teff+  & $\Delta$ \logg+ & $\Delta$ \Vt+ & tot+\\
\hline
42	&SrI	&0.03 &0.04  &0.04	&0.06	&0.03	&0.05  &0.03	&0.07	 \\ 
42	&ZrI	&0.13 &0.01  &0.00	&0.13	&0.14	&0.01  &0.00	&0.14	 \\ 
42	&MoI	&0.14 &0.01  &0.00	&0.14	&0.15	&0.02  &0.00	&0.15	 \\ 
44	&RuI	&0.14 &0.02  &0.00	&0.14	&0.15	&0.03  &0.00	&0.15	 \\ 
\hline                             
\end{tabular}
\end{table*}

The resulting errors in determining the Mo and Ru abundances due to the uncertainties in the atmospheric parameters do not exceed 0.15 dex.
The stellar parameters and the obtained [Sr/Fe], [Zr/Fe], [Mo/Fe], and [Ru/Fe] abundances are provided in machine-readable
format at the Centre de Données astronomiques de Strasbourg
(CDS) in Table A1.

\subsection{Impact of the departures from LTE }

Notice that our abundance determinations are performed in the LTE approximation. Of the four elements considered here, our results for Zr are potentially the
most prone to systematic errors from this approximation. The nonlocal thermodynamic equilibrium (NLTE) study of \cite{velichko:10}  found that the minority neutral species of Zr is susceptible to over-ionization, with corrections on the order of $+0.30$ dex in red giant branch stars for the two strong Zr I lines considered in that work. However, it is unclear what corrections would be found for the weak Zr I lines
considered in our work. Furthermore, while their study employs a large model atom, it should be kept in mind that it precedes recent advances in our understanding of inelastic hydrogen collisions \citep[e.g.,][]{caliskan:25}; therefore, the quantitative results of that study may change significantly.
In contrast, our Sr abundances are probably much less prone to systematic errors due to the LTE approximation. This is because we used ionized Sr II lines, which,
as shown by \cite{bergemann:12}, are less sensitive to LTE departures. For example, the corrections are zero for the 4077 Sr II line for stellar parameters similar to those of our representative stars 
\citep[see Table 4,][]{bergemann:12}.
For Mo and Ru, there is no specific literature; however, for a qualitative discussion one may consider NLTE results for lines from analogous species. For
the Mo I lines at 5506.49 and 5533.03 ($\mathrm{4d^{5}(^{6}S)5s a ^{5}S_{2}}$ --
$\mathrm{4d^{5}(^{6}S)5p z ^{5}P^{\mathrm{o}}_{3,2}}$), close analogies are the Cr I lines at 5208.41 and 5206.02 ($\mathrm{3d^{5}(^{6}S)4s a ^{5}S_{2}}$ --
$\mathrm{3d^{5}(^{6}S)4p z ^{5}P^{\mathrm{o}}_{3,2}}$). The MPIA NLTE database \citep{kovalev:19}, which employs the model atom of \cite{bergemann:10}, provides corrections of around $+0.15$ dex for these two lines and the two
representative stars. This is of similar order to the uncertainties propagated from the stellar parameters (Table 3) and is therefore potentially significant.
For the Ru I lines at 4584.44 ($\mathrm{4d^{7}(a ^{4}F)5s a ^{3}F_{3}}$ --
$\mathrm{4d^{7}(a ^{4}F)5p z ^{3}G^{\mathrm{o}}_{4}}$) and 4757.76
($\mathrm{4d^{6}5s^{2} a ^{5}D_{1}}$ -- $\mathrm{4d^{6}5s(a ^{5}D)5p z
^{7}F^{\mathrm{o}}_{1}}$), possible analogies are the Fe I lines at 4307.90
($\mathrm{3d^{7}(^{4}F)4s a ^{3}F_{3}}$ -- $\mathrm{3d^{7}(^{4}F)4p z
^{3}G^{\mathrm{o}}_{4}}$) and 4471.68 ($\mathrm{3d^{6}4s^{2} a ^{5}D_{1}}$ --
$\mathrm{3d^{6}(^{5}D)4s4p(^{3}P^{\mathrm{o}}) z ^{7}F^{\mathrm{o}}_{1}}$).  For
the first line, the 1D NLTE data of \cite{amarsi:16} suggest slightly negative corrections of around $-0.05$ dex for the two representative stars.
The second line is not included in the dataset of \cite{amarsi:16}; however, lines from the same multiplet are present. These have corrections of around $-0.03$ dex on average for the two representative stars. These are much less than the uncertainties propagated from the stellar parameters (Table 3). 
In summary, we suspect that, in the present study, the results for Zr would be most affected by departures from LTE, followed by Mo; whereas Ru and Sr may have
insignificant effects relative to the other uncertainties. Detailed NLTE modeling for these elements based on modern atomic data is crucial to verify and quantify this claim.

\section{Results and abundance signatures}
\label{sec: results}
 
\begin{figure}
\begin{tabular}{c}
\includegraphics[width=7.6cm]{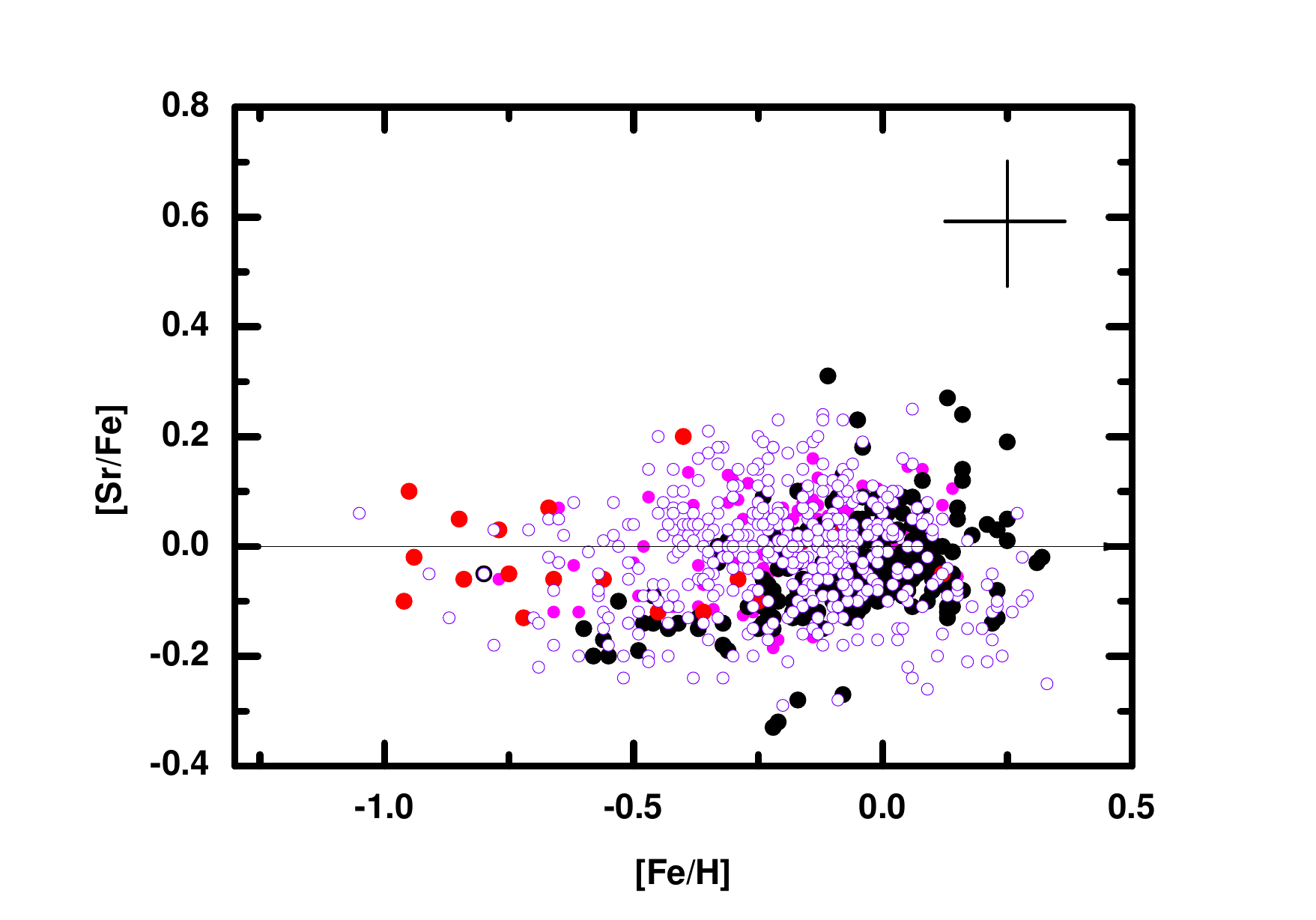}\\
\includegraphics[width=7.6cm]{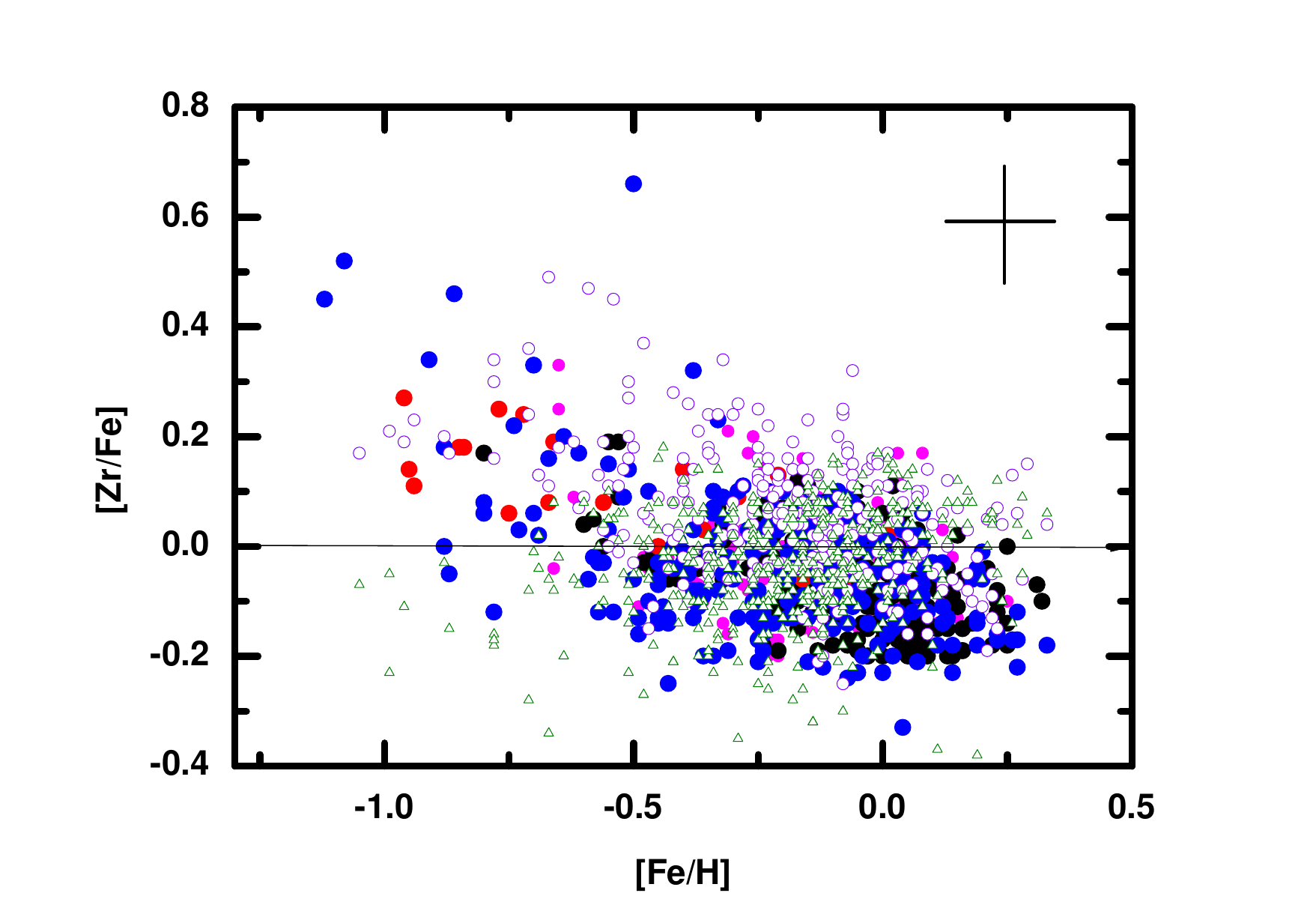}\\
\includegraphics[width=7.6cm]{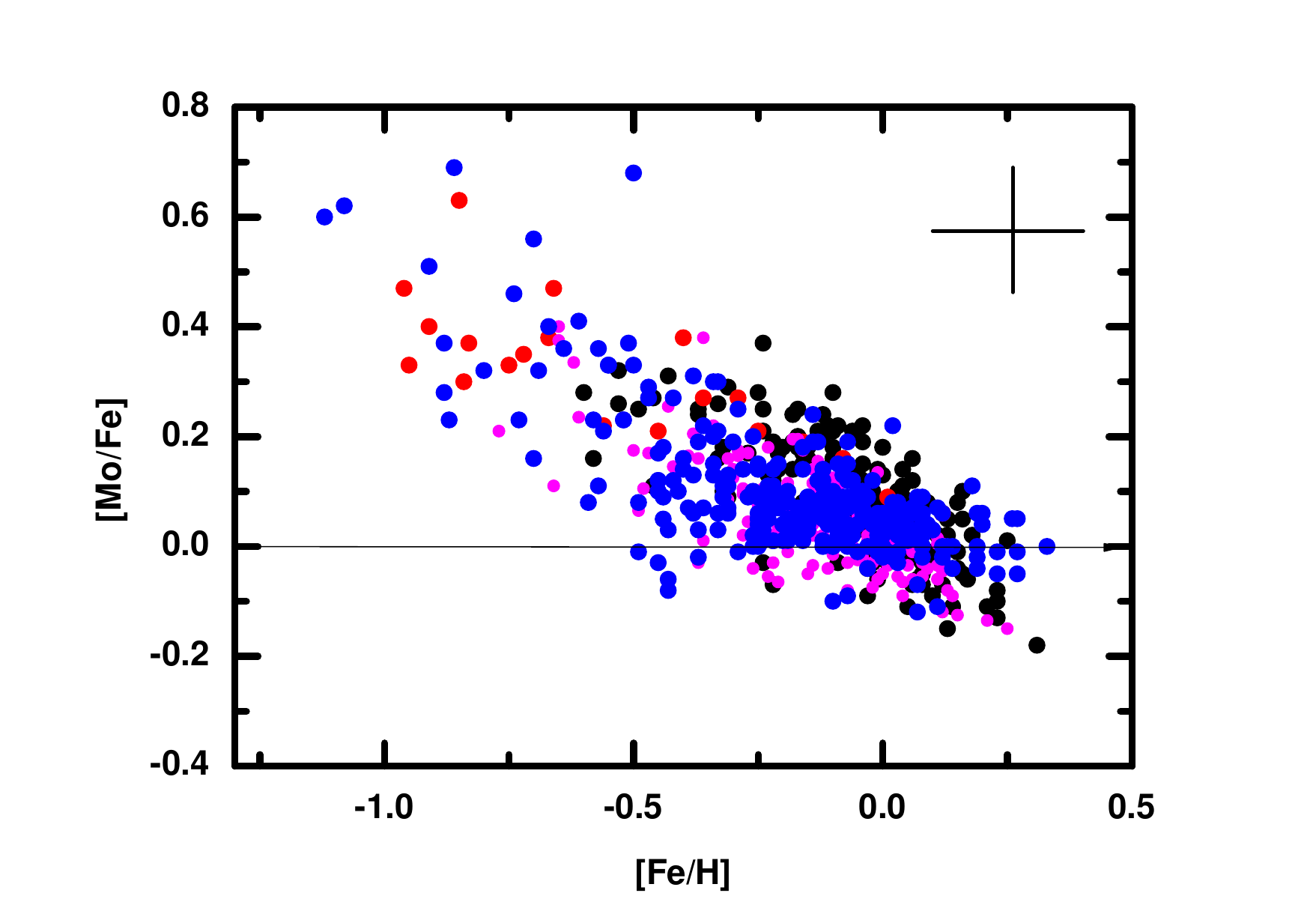}\\
\includegraphics[width=7.6cm]{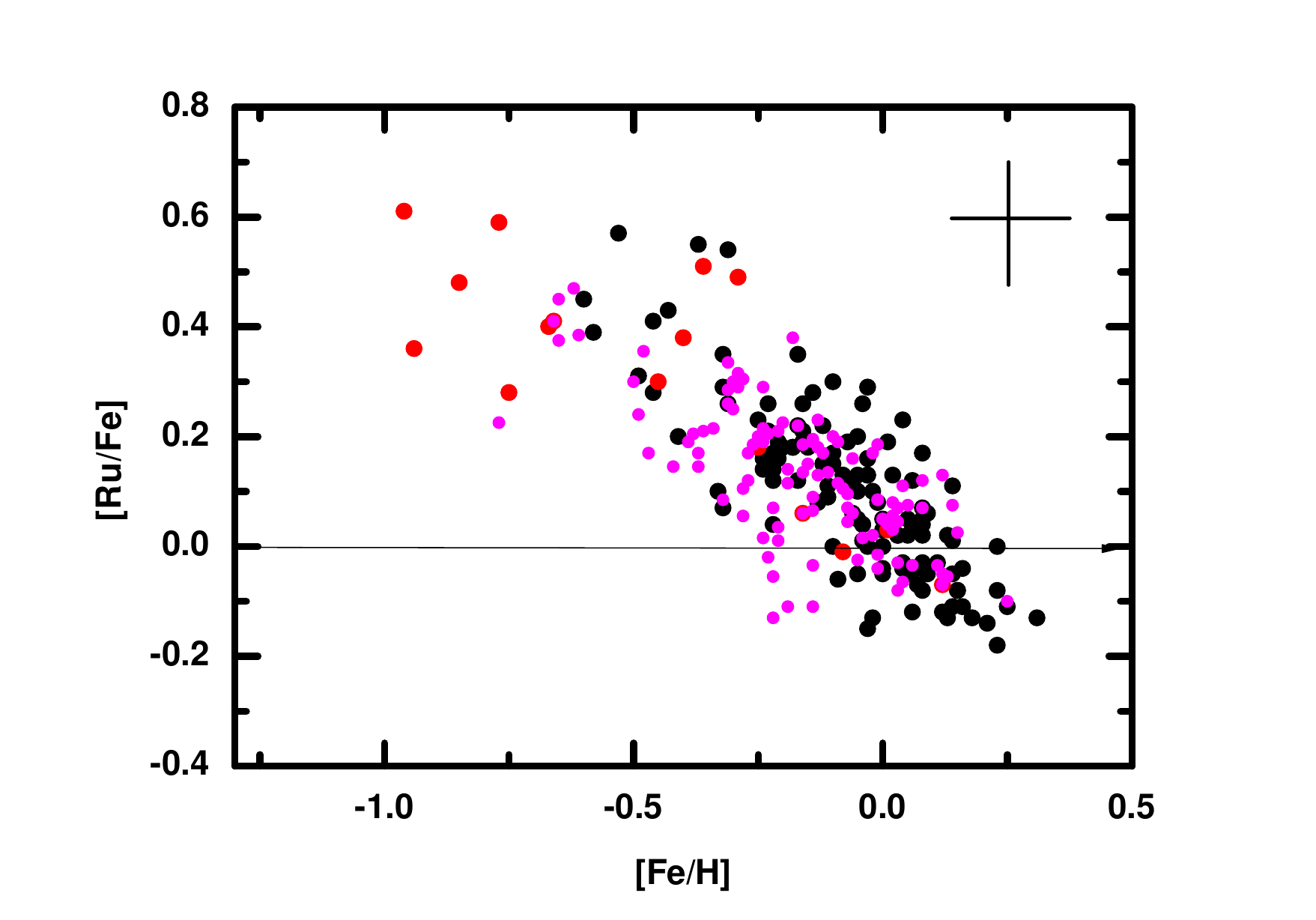}\\
\end{tabular}
\caption{[Sr/Fe], [Zr/Fe], [Mo/Fe], and [Ru/Fe] with respect to [Fe/H] given for the giant stars in our sample (magenta points), the dwarf stars by \cite{mishenina:19a, mishenina:19b, mishenina:13} (thin disk stars as black points, and thick disk stars as red points). The data from \cite{forsberg:19} (Zr) and \cite{forsberg:22} (Mo) are shown as blue points, while those from \cite{tautvai:21} -- Sr and Zr I -- are shown as open purple circles, and Zr II as open green triangles.}
\label{elfe_feh}
\end{figure}

We report in Figure \ref{elfe_feh}, the [Sr/Fe], [Zr/Fe], [Mo/Fe], and [Ru/Fe] with respect to [Fe/H], compared with the stellar sample of dwarf stars from our previous studies -- Mo and Ru \citep[][]{mishenina:19a}, Sr \citep[][]{mishenina:19b}, Zr \citep[][]{mishenina:13} -- obtained for the same set of stars with the same stellar parameters. In these figures, we also present data from \cite{forsberg:19} (Zr) and \cite{forsberg:22} (Mo), as well as data from \cite{tautvai:21} (Sr) and Zr abundances determined from both Zr I and Zr II lines, since they used both sets of lines in their analysis. The abundance trends are generally consistent between the two sets of stars: the Sr trend with [Fe/H] is roughly flat, while the other three elements tend to decrease with increasing metallicity, reaching subsolar ratios for stars with metallicities higher than solar.   

\begin{figure}
\includegraphics[width=8cm]{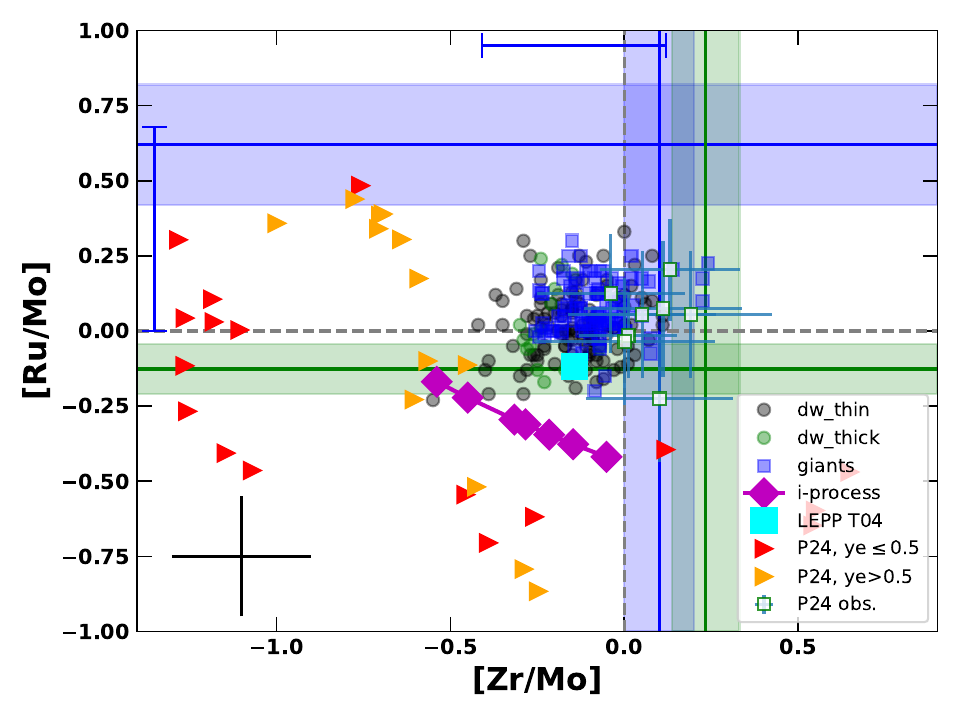}\\
\caption{[Ru/Mo] for the stellar sample presented in this work, shown with respect to [Zr/Mo]. For comparison, the metal-poor stars from \cite{honda:07,hansen:12,Hansen:14,Aoki:17} and \cite{psaltis:24} are shown as white squares with green contours. The ratios from the $s$-process GCE simulations by \cite{bisterzo:14} (green line and error range) and from the r-II star CS 22892-052 \citep[][]{sneden:03} (green line and error range) are included as references. The production range derived from a collection of r-II stars is indicated by the blue horizontal and vertical error bars for [Zr/Mo] and [Ru/Mo], respectively (see the text for details). We also report the LEPP prediction by \cite{travaglio:04} (large cyan square), the ratios from different neutrino-driven ejecta conditions in \cite{psaltis:24} with $Y_{\rm e}$ $>$ 0.5 (orange triangles, $\nu p$-process) and with $Y_{\rm e}$ $<$ 0.5 (red squares, weak \textit{r}-process), along with $i$-process predictions for mild neutron exposures, characterized by production peaking in the Zr-Ru region, and no relevant production at the Ba peak (magenta-lined squares; see text for more details). The typical observation error for our stars is given in the lower left corner of the plot (black error bar). }
\label{fig:zrmo_vs_rumo}
\end{figure}

In order to study the abundance signatures for the stellar sample independently of [Fe/H], in Figure \ref{fig:zrmo_vs_rumo} the measured [Zr/Mo] ratios with respect to [Ru/Mo] are shown. There are no stars showing lower [Ru/Mo] or higher [Zr/Mo] than those predicted by the $s$-process in GCE model \citep[][]{bisterzo:14} or observed in $r$-process stars. In particular, for the elements discussed here, the derivation of a classical $r$-process residual for the solar system is extremely uncertain \citep[][]{arlandini:99,travaglio:04, bisterzo:14, prantzos:20}.  
The r-II star CS 22892-052 exhibits an abundance pattern compatible with the solar $r$-process residual for elements heavier than Ba \citep[][]{sneden:03, sneden:08}. 
However, \cite{travaglio:04} found that this consistency does not hold for elements between Sr and Ba, leading to the definition of the additional $s$-process component LEPP (see discussion in the Introduction), 
to fully reproduce the solar composition. The predicted LEPP signature by \cite{travaglio:04} is shown in Figure \ref{fig:zrmo_vs_rumo}.
The LEPP component was originally associated with the abundances observed in several metal-poor stars, exhibiting high production at the Sr peak and marginal or no production of the lanthanides, i.e., it was not produced together with $r$-process elements \citep[][]{montes:07}. Some of these Sr-rich metal-poor stars, taken from \cite{psaltis:24}, are also reported for comparison, together with different conditions of the neutrino-driven ejecta with varying electron fraction ($Y_e$) provided in the same work. We also report the typical range of ratios obtained from the $i$-process conditions \citep[for more details about the simulations, see e.g.,][]{bertolli:13, mishenina:15, roederer:16, roederer:22, ji:24}, calculated for neutron exposures yielding an efficient production of the Zr-Ru mass region but without producing Ba, since this additional contribution to Ba is not observed by definition in solar abundances \citep[e.g.,][]{mishenina:15, baratella:21} or in metal-poor LEPP stars \citep[][and references therein]{montes:07, sitnova:25}. 

Within observational uncertainties, our stars and the LEPP stars by \cite{psaltis:24} show similar abundance ratios in Figure \ref{fig:zrmo_vs_rumo}. The LEPP component derived from \citep[][]{travaglio:04} is compatible with the stellar ratios, while the $i$-process component spans the [Zr/Mo] range, but shows about 0.2 dex lower [Ru/Mo]. For the [Zr/Mo] range covered by our stars, neutrino wind trajectories with varying electron fraction y$_{\rm e}$ tend to have subsolar [Ru/Mo] values, while for the [Ru/Mo] range covered by our stars, they have subsolar [Zr/Mo] values. 

\section{Further discussion: Tracing the origin of Zr, Mo, Ru, and their nuclear uncertainties}
\label{sec: discussion}

From the results in Figure \ref{fig:zrmo_vs_rumo}, an interesting question would be whether it is possible to identify the need for an anomalous nucleosynthesis contribution, in addition to the $s$-process and the $r$-process signatures, from [Zr/Mo] and [Ru/Mo]. Such a result would also be consistent with \cite{forsberg:22} and \cite{vanderswaelmen:23}, where at least an additional Mo nucleosynthesis component was identified by comparing Mo abundances with Eu, in stars from both the Milky Way disk and the galactic Bulge. However, while these previous studies have used the $r$-process element Eu as the main reference to infer the production of Mo, or GCE of multiple $r$-process and $s$-process elements, as in \cite{molero:23}, in this work we adopt a complementary approach, focusing on the relative production signatures of the nearby elements Zr, Mo, and Ru. 

Let us assume that CS 22892-052 carries abundance signatures representative of the $r$-process production in the Zr-Ru region. Its [Ru/Mo] in Figure \ref{fig:zrmo_vs_rumo} is larger than all MW disk stars, and the [Zr/Mo] is similar to the $s$-process GCE ratio.
Hence, in this scenario the [Ru/Mo] ratio for all our stars could be explained by a combination of $s$-process and $r$-process contributions, without the need for other nucleosynthesis components. However, we would need one or more additional processes that eject abundances with [Zr/Mo] lower than solar. Such a signature seems possible for either the $i$-process, the solar LEPP, or the neutrino-driven ejecta, which exhibit the lowest [Zr/Mo]. Such a result would generally be aligned with \cite{forsberg:22} and \cite{vanderswaelmen:23}, although at present we cannot clearly identify if there are one or more responsible nucleosynthesis processes. 

However, let us now consider an extended sample of r-II stars that are highly $r$-process enriched with [Eu/Fe]$>$1 and follow a [Ba/Eu] $<$ 0 \citep[e.g.,][]{farouqi:22}. In  Figure  \ref{fig:zrmo_vs_rumo} we report the production range for additional r-II stars with measurements available for all four elements: 2MASS J22132050–5137385 \citep[][]{roederer:24}, 2MASS J00512646-1053170 \citep[][]{shah:24}, RAVE J183013.5-455510 \citep[][]{placco:20}, CD-45 3283, HD 3567, and HD 120559 \citep[][]{hansen:12,Hansen:14}, HD 222925 and BD +17°3248 \citep[][]{roederer:22}, and HE 1523-0901 \citep[][]{mishenina:24}. We indeed find great variation in these ratios. The [Zr/Mo] range extends from the mildly super-solar values observed in CS 22892-052 down to -0.45 dex. For [Ru/Mo], the super-solar values observed for CS 22892-052 are extended downward by almost 0.7 dex, covering the full range down to solar. With such variation, the entire [Zr/Mo] range observed in our stars can be explained by a combination of $s$-process and $r$-process nucleosynthesis. While it is established that r-II stars exhibit significant spread in the production of elements in the Zr-Ru region relative to Eu \citep[e.g.,][]{sneden:08}, it has been shown that some correlated production in this mass region should be present relative to the $r$-process peak element Te \citep[][]{roederer:22}. 

In the future, a specific study should be conducted on r-II stars to verify whether the current scatter in [Ru/Mo] and [Zr/Mo] represents a true nucleosynthesis signature, or results from errors introduced by different spectroscopic analyses. This would help us use these elements as diagnostics to identify the need for additional nucleosynthesis components in the MW disk beyond the classical $s$-process and $r$-process contributions. 

In Figure \ref{fig:zrmo_vs_rumo}, neither the $i$-process calculations nor the neutrino-driven ejecta components show the impact of the present nuclear uncertainties on the simulations. 
For the $i$-process, estimations of the impact of relevant neutron-capture rate uncertainties are available for the Sr, Y, and Zr peak by \cite{denissenkov:18}, and for a more extended region including Mo and Ru by \cite{mckay:20}. In these results, variations of 1 dex or larger is shown for each element. \cite{goriely:21} also estimated the impact of nuclear uncertainties on $i$-process calculations in asymptotic giant branch (AGB) stars at low metallicity, reporting typical errors on the order of 0.4 dex. However, the impact on specific elemental ratios (in this case [Zr/Mo] and [Ru/Mo]) should be explored in detail. At the moment it is not clear what the relevant $i$-process stellar sites (if any) contributing to the chemical abundances in the MW disk are \citep[e.g.,][]{cote:18a, choplin:24}.
For neutrino-driven ejecta conditions (weak $r$-process and $\nu p$-process), the impact of thermonuclear reaction rate uncertainties has been explored in several studies~\citep{nishimura:19,bliss:20,psaltis:22}. Because these nucleosynthesis pathways lie close to the valley of stability, many key reactions are within reach of current and upcoming radioactive ion beam facilities~\citep{dilling:13,brown:25}. Currently, nuclear physics uncertainties often exceed the precision of stellar abundance measurements~\citep{psaltis:22}, hindering efforts to disentangle astrophysical conditions from nuclear effects. Improved constraints on these rates will directly strengthen the link between models and observations. Based on previous studies mentioned above, we can estimate errors on the order of 0.5 dex for Zr, Mo, and Ru. 

In Figure \ref{fig:zrmo_vs_rumo} we show elemental ratios measured in $r$-process stars, rather than directly from $r$-process simulations, where nuclear uncertainties have a strong impact on the calculated yields. 
Among key nuclear inputs are $\beta$-decay half-lives ($T_{!1/2}$) and one- and two-neutron emission probabilities (P1n and P2n). 
Recently, \cite{Tolosa:25} demonstrated a significant impact on the calculated abundances of elements up to Mo, showing an increase in Y and Mo by $\sim$ 70\%, and in Zr by $\sim$ 50\%, when comparing the standard JINA REACLIB database \citep{Cyburt:10} with the same database updated with new experimental values. 
A comparison with the metal-poor star from \cite{honda:07} is also presented in Tolosa's work, showing a clear overestimation of Mo relative to the star from \cite{honda:07}. At present, the dispersion in the data for the r-II stars shown in Figure \ref{fig:zrmo_vs_rumo} does not likely provide strong constrains for $r$-process calculations. However, by reducing observational uncertainties in the abundance determination of these stars, important constraints could also be derived for nuclear uncertainties. New nuclear data on $\beta$-decay half-lives and Pn values are needed to deepen these studies, and such data are expected in the near future \citep[][]{LNL:23}.

\section{Summary and conclusions}
\label{sec: conclusions}

We present a new analysis of the abundances of molybdenum (Mo, Z = 42) and ruthenium (Ru, Z = 44) in the Galactic disk (–1.0 < [Fe/H] < +0.3), based on 
a new sample of giant stars. New determinations of Mo and Ru abundances 
were performed using high-resolution spectra (R = 42,000) with signal-to-noise ratios (S/N) exceeding 100, obtained with the ELODIE spectrographs (Observatoire de Haute-Provence, France). Mo abundances were derived from Mo I lines at 5506 and 5533 \AA~ for 154 stars, while Ru abundances were determined from Ru I lines at 4584 and 4757 \AA~ for 110 stars, using the LTE approach with an accuracy of 0.15 dex. For the majority of the target giants, Mo and Ru abundances were determined for the first time. Additionally, the abundances of strontium (Sr) and zirconium
(Zr) were also measured.
In addition to the giants, we used data from the stellar sample of dwarf stars from our previous papers \citep{mishenina:19a, mishenina:19b, mishenina:13} in our analysis. The abundance trends ([Sr/Fe], [Zr/Fe], [Mo/Fe], and [Ru/Fe] with respect to [Fe/H]) are consistent between the two sets of stars.

We studied the abundance signatures for the stellar sample independently from [Fe/H], using the measured [Zr/Mo] ratios with respect to [Ru/Mo]. 
While several nucleosynthesis processes have been proposed as a possible source of Mo and Ru, it is uncertain at present if the $s$-process and the $r$-process contributions could fully explain observations in the MW disk (and in the solar system). Based on the observed range of [Zr/Mo] and [Ru/Mo], a combination of the $s$-process and the abundance scatter seen in r-II stars could, in principle, allow us to reproduce the data. In particular, we have shown that it is crucial to determine whether the current [Zr/Mo] scatter observed in r-II stars (on the order of $\sim$0.5 dex) is real or mostly due to observational uncertainties. In this second case, explaining the range of [Zr/Mo] observed in the MW disk would require additional nucleosynthesis sources, such as the $i$-process and/or neutrino-driven ejecta components that are either proton-rich or neutron-rich, which eject abundances with [Zr/Mo] lower than solar.  

Finally, we emphasize that defining NLTE corrections for Mo and Ru abundance measurements is essential, as these remain currently unknown.

\section{Data availability}

Table A1 is only available in electronic form at the CDS via anonymous ftp to cdsarc.u-strasbg.fr (130.79.128.5) or via http://cdsweb.u-strasbg.fr/cgi-bin/qcat?J/A+A/.
    
\begin{acknowledgements} 
MP acknowledges significant support to NuGrid from NSF grant PHY-1430152 (JINA Center for the Evolution of the Elements) and STFC (through the University of Hull's Consolidated Grant ST/R000840/1), and access to {\sc viper}, the University of Hull High Performance Computing Facility. MP acknowledges the support from the "Lendulet-2023" Program of the Hungarian Academy of Sciences (LP2023-10, Hungary) and from the NKFI via K-project 138031 (Hungary).
 MP acknowledges support from the ERC Consolidator Grant (Hungary) funding scheme (project RADIOSTAR, G.A. n. 724560), from the ERC Synergy Grant Programme (Geoastronomy, grant agreement number 101166936, Germany) and from the National Science Foundation (USA) under grant No. PHY-1430152 (JINA Center for the Evolution of the Elements). This article is based upon work from the ChETEC COST Action (CA16117), supported by COST (European Cooperation in Science and Technology).
We thank the ChETEC-INFRA project funded from the European Union's Horizon 2020 research and innovation programme (grant agreement No 101008324), and the IReNA network supported by NSF AccelNet. MP also thank the UK network BRIDGCE. AMA acknowledges support from the Swedish Research Council (VR 2020-03940) and the Crafoord Foundation via the Royal Swedish Academy of Sciences (CR 2024-0015). TM is grateful to the Laboratoire d'Astrophysique de l'Universite de Bordeaux for their kind hospitality. 
We acknowledge support from Grant PID2023-147569NB-C21 funded by MICIU/AEI/10.13039/501100011033 Spain and ERDF, UE. The authors would like to thank the anonymous referee for helpful and detailed comments on the manuscript.
\end{acknowledgements}

\bibliography{molybdenum}

\end{document}